%% file: main.tex
\documentclass{itor}

\usepackage{natbib}%
\usepackage[figuresright]{rotating}

\usepackage{lscape, comment}
\usepackage{url}
\usepackage{algorithmic}
\usepackage{amsmath,amssymb,amsfonts}
\usepackage{algorithmic}
\usepackage{graphicx, bbm}
\usepackage{textcomp, nccmath}
\usepackage{xcolor}
\usepackage{amsmath}
\usepackage{algorithm}
\usepackage{caption}
\usepackage{subcaption}

\usepackage{amsmath,amsthm}

\theoremstyle{definition}

\newcommand{\blue}[1]{\textcolor{black}{#1}}
\newcommand{\red}[1]{\textcolor{black}{#1}}
\theoremstyle{remark}

\newcommand{\algo}{PASA}

\jname{International Transactions in Operational Research}
\jvol{XX}
\jyear{20XX}
\doi{xx.xxxx/itor.xxxxx}

\begin{document}



\title{\algo{}: \emph{A Priori} Adaptive Splitting Algorithm for the Split Delivery Vehicle Routing Problem}

\author[Running Author]{Nariman Torkzaban\affmark{1}\affmark{$\ast$}, Anousheh Gholami\affmark{1}\affmark{$\ast$}, John S. Baras\affmark{1}, and Bruce Golden\affmark{2}}

\affil{\affmark{1}Department of Electrical and Computer Engineering, University of Maryland, College Park, MD, USA}
\affil{\affmark{2}Robert H. Smith School of Business, University of Maryland, College Park, MD, USA}
\email{narimant@umd.edu [Torkzaban]; anousheh@umd.edu [Gholami];\\ baras@umd.edu [Baras]; bgolden@umd.edu [Golden] }

\thanks{\affmark{$\ast$} The identified authors contributed equally to this paper.}


\begin{abstract}
The split delivery vehicle routing problem (SDVRP) is a relaxed variant of the capacitated vehicle routing problem (CVRP) where the restriction that each customer is visited precisely once is removed. Compared with CVRP, the SDVRP allows a reduction in the cost of the routes traveled by vehicles. The exact methods to solve the SDVRP are computationally expensive. Moreover, the complexity and difficult implementation of the state-of-the-art heuristic approaches hinder their application in real-life scenarios of the SDVRP. In this paper, we propose an easily understandable and effective approach to solve the SDVPR based on an \emph{a priori} adaptive splitting algorithm (\algo{}). The idea of \emph{a priori} split strategy was first introduced in \cite{base}. In this approach, the demand of the customers is split into smaller values using a fixed splitting rule in advance. Consequently, the original SDVRP instance is converted to a CVRP instance which is solved using an existing CVRP solver. While the proposed \emph{a priori} splitting rule in \cite{base} is fixed for all customers regardless of their demand and location, we suggest an adaptive splitting rule that takes into account the distance of the customers to the depot and their demand values. Our experiments show that \algo{} can generate solutions comparable to the state-of-the-art but much faster. Furthermore, our algorithm outperforms the fixed \emph{a priori} splitting rule proposed in \cite{base}.
\end{abstract}


\maketitle

\input{introduction}

\input{SDVRP}
\input{solution}

\input{evaluation}
\clearpage\section{Conclusions}
\label{sec:conclusion}
We propose an efficient and effective heuristic algorithm for the SDVRP based on a two-step procedure of splitting the demand of customers into smaller values and solving the generated network assuming it to be a CVRP instance. Our approach focuses on enhancing the splitting rule proposed by \cite{base} by considering customer locations and demand values. Our proposed adaptive splitting algorithm incorporates clustering customers based on their location and choosing a specific splitting rule for each cluster. It is shown to outperform the fixed splitting rule of the approach adopted in \cite{base}, for all four benchmark datasets on average, both with respect to performance and computational effort. 

\section*{Acknowledgments}
The research of Torkzaban, Gholami, and Baras was partially supported by ONR grant N00014-17-1-2622, and by a grant from Lockheed Martin Chair in Systems Engineering.


\nocite{*}

\bibliographystyle{itor}
\bibliography{itor}




\end{document}

%% file: introduction.tex
\section{Introduction}
\label{sec:introduction}
The split delivery vehicle routing problem (SDVRP) introduced by \cite{dror1989savings} is a relaxation of the traditional capacitated vehicle routing problem (CVRP) where each customer can be visited more than once. In both CVRP and SDVRP, a number of identical vehicles having limited capacity serve a set of customers with given demands. The vehicles depart a depot and return to the depot after visiting customers. The pairwise travel costs between customers and between customers and the depot are given. The objective is to minimize the total travel cost of the vehicles. The difference between CVRP and SDVRP is that in contrast to CVRP where each customer is required to be visited once and by only one vehicle, SDVRP allows multiple visits to a customer. Therefore, the demand of a customer may be split among multiple vehicles. \cite{bounds} show that splitting the customer demands potentially reduces the total cost by up to $50\%$. Further, they show that there always exists an optimal solution to the SDVRP in which there is no k-split cycle and no two routes have more than one common customer. These findings justify the essence of formulating and solving the SDVRP.

The SDVRP can be formulated as a mixed integer linear program (MILP) \citep{ArchettiCG}. To tackle the time complexity of the MILP, various common optimization methods as well as meta-heuristics have been proposed. \cite{MinMax19*} introduced a max-min clustering before solving the optimization problem. Tabu search is used in several papers authored by \cite{wang17*, xia18, yang18}. \cite{wang17*} proposed an approach relying on simulated annealing to solve the SDVRP. Column generation is another optimization technique that has been applied to the SDVRP by \cite{ArchettiCG, JinCG}. \cite{chess19} applied the cutting plane method to tackle the time complexity of the SDVRP.
Recently, \cite{ILS-MIP} introduced a metaheuristic approach incorporating several mathematical programming components within an iterative local search framework. We refer to this method as ILS-MIP in the rest of the paper. ILS-MIP starts from an initial set of solutions obtained using the I1 Solomon heuristic followed by a perturbation step (to escape local optima), a local search step using classical neighborhood search heuristics, and several steps based on hybrid components. The hybrid components are called only if the incumbent solution of the search is not improved after a number of steps. The hybrid components encompass a mixed integer program (MIP)-based improvement heuristic performing two operations of inserting splits or removing potential unnecessary splits. Another used hybrid component is based on converting the current SDVRP solution to a CVRP instance and using a hybrid genetic search (HGS) framework \citep{HGS} specialized to solve CVRP. An MIP model is also used in the third component of the hybrid step where a residual problem is constructed by removing the edges that are not used in any solution of the current set of solutions. The residual problem is then solved using a MIP solver. Finally, the fourth component uses the branch-and-cut (BC) framework proposed by \cite{BC}. While ILS-MIP is a powerful and effective methodology, it is not practical in real-world problems due to its complexity and huge run time.

Although the state-of-the-art algorithms for the SDVRP are capable of producing very high-quality solutions, they are not easily understandable, have high computational complexity, and, therefore, are very difficult to implement in real-world scenarios. To this end, \cite{base} proposed an efficient algorithm that decomposes the SDVRP into two sub-problems. First, a demand-splitting rule is applied to all customers. By doing so, the demand of each customer is split into smaller demands, each representing a new customer located in the same position as the original customer. As a result, a new problem instance is generated with an increased number of customers. Second, the new problem is assumed to be an instance of the CVRP which is solved using existing powerful CVRP solvers. The transformation of the SDVRP to a CVRP not only facilitates the understandability and implementation of SDVRP but also enables leveraging the rich literature on the CVRP. The obtained solution for the resulting CVRP is then translated back to the original SDVRP. 

In the method proposed by \cite{base}, a fixed splitting rule that is inspired by the US coin denominations is applied to all customer demands and the VRPH solver introduced by \cite{VRPH} is used to solve the resulting CVRP instance. We refer to this algorithm as VRPHAS in the rest of this paper. The numerical evaluation of VRPHAS illustrates its ability to produce acceptable sub-optimal solutions in much less time compared to the state-of-the-art heuristic algorithms. However, the adoption of the introduced splitting rule is not well justified. Instead of applying a fixed rule to all customers, a splitting rule that is adaptive, based on the specific characteristics of a customer can potentially result in an improved solution for the SDVRP. In this paper, we study the problem of defining a good splitting rule for each customer. To this end, we introduce an \emph{a priori} adaptive splitting algorithm (PASA). In combination with the VRPH solver, \algo{} results in an improved performance in terms of the optimality gap without significantly increasing the computational complexity of VRPHAS. \algo{} achieves this by taking into account the distance between the customers and the depot, the value of the customers' demand along with the vehicles' capacity. Similar to VRPHAS, \algo{} can be used with any CVRP solver. Thus, it provides a method to solve the SDVRP that is easily understandable, can be implemented simply, and generates high-quality solutions very quickly. 

The rest of the paper is organized as follows. In Section \ref{sec:problem}, we provide a formal definition of the SDVRP. Our proposed solution is explained in Section \ref{sec:solution}. Section \ref{sec:evaluation} presents the numerical results. Finally, the conclusion is discussed in Section \ref{sec:conclusion}.

%% file: SDVRP.tex
\section{SDVRP Definition}
\label{sec:problem}
Let $G=(V,E)$ denote an undirected and weighted complete graph representing the network of customers and a depot. The vertex set $V=\{0,\dots ,n\}$ represents the depot denoted by the vertex $0$ and $n$ customers indexed as $1,\dots,n$. A non-negative weight $c_{ij}$ is associated with each edge of the graph $(i,j)\in E$ that stands for the cost of traveling between customers $i$ and $j$. Moreover, each customer $i\in V$ has a positive demand, denoted by $d_i$. A set of $M$ vehicles is available to serve the customers' demands. Each vehicle has a limited capacity denoted by $Q$. The objective of the SDVRP is to find a route for each vehicle such that the demands of all customers are satisfied with the minimum total cost. In this paper, we assume that the cost of traveling between customers $i, j\in V$ is their Euclidean distance denoted by the function $dist$, i.e. $c_{ij} = dist(i, j)$.

%% file: solution.tex
\section{Proposed Heuristic Algorithm for the SDVRP}
\label{sec:solution}

In this section, we first describe the idea of \emph{a priori} splitting rule proposed by \cite{base} and then discuss the core algorithmic idea of \algo{} based on two motivational observations. We then formalize \algo{} in detail. 
\subsection{\emph{A priori} splitting rule}
The proposed splitting rule in \cite{base} is inspired by the US coin denominations. The authors propose two splitting options $20/10/5/1$, and $25/10/5/1$. The first option replaces customer $i$, with the demand of $d_i$ by $m^i_{20}$, $m^i_{10}$, $m^i_{5}$ and $m^i_{1}$ number of customers with the demand values of $0.2Q$, $0.1Q$, $0.05Q$ and $0.01Q$, respectively. Hence, we have:
\begin{equation}
    m^i_{20} = max\{m\in \mathbb{Z}^+ \cup \{0\}: 0.2 Q m \leq d_i\};
\end{equation}
\begin{equation}
    m^i_{10} = max\{m\in \mathbb{Z}^+ \cup \{0\}: 0.1 Q m \leq d_i - 0.2 Q m^i_{20}\};
\end{equation}
\begin{equation}
    m^i_{5} = max\{m\in \mathbb{Z}^+ \cup \{0\}: 0.05 Q m \leq d_i - 0.2 Q m^i_{20} - 0.1 Q m^i_{10}\};
\end{equation}
\begin{equation}
    m^i_{1} = max\{m\in \mathbb{Z}^+ \cup \{0\}: 0.01 Q m \leq d_i - 0.2 Q m^i_{20} - 0.1 Q m^i_{10} - 0.05 Q m^i_{5}\}.
\end{equation}
A similar breakdown can be given for the second splitting rule option.
Using the splitting rule $20/10/5/1$, the demand of customer $i$ is split into $m^i = m^i_{20}+m^i_{10}+m^i_{5}+m^i_{1}$ smaller values. The customer $i$ is then replaced by $m^i$ customers located at the same position as customer $i$ and with the demand values resulting from $(1)-(4)$. After applying the splitting rule to all customers, a new graph is constructed with $m = \sum_{i=1}^n m^i$ vertices (customers). The new graph is considered as an instance of the CVRP and the VRPH solver is used to solve it. The resulting solution is also a solution to the original SDVRP instance. In the next section, we justify the essence of the adaptation of the splitting rule and propose the \algo{} algorithm.

\subsection{\emph{A priori} Adaptive Splitting Algorithm (\algo{})}
The \emph{a priori} splitting rule described by equations $\text{(1)-(4)}$ is a \emph{fixed} rule, i.e., a single splitting rule is applied to all customers regardless of their demand and location. However, the information regarding the customers' demand and location can be leveraged in the vehicle routing decisions towards reducing the total travel cost. We propose an adaptive splitting rule based on the location and demand information that is shown to effectively improve the solution. In the following, we first present the motivation behind the main building blocks of our proposed splitting rule. We then explain the \algo{} algorithm.
\subsubsection{Motivation}
\begin{figure}[h!]
     \centering
     \begin{subfigure}[b]{0.42\textwidth}
         \centering
         \includegraphics[width=1\textwidth]{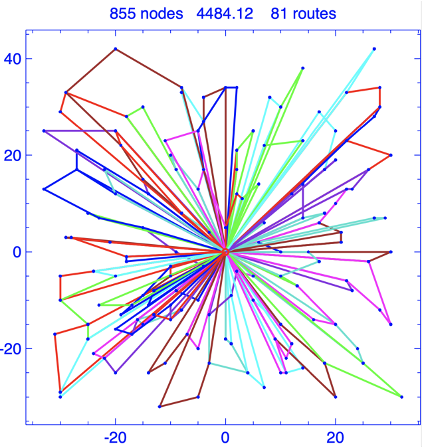}
         \caption{20/10/5/1 rule, \textbf{231.91 s}}
         \label{fig:obs1-1}
     \end{subfigure}
     \begin{subfigure}[b]{0.42\textwidth}
         \centering
         \includegraphics[width=\textwidth]{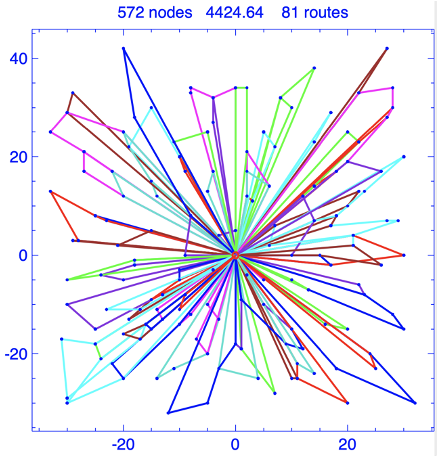}
         \caption{128/64/32/16/8/4/2 rule, \textbf{86.67 s}}
         \label{fig:obs1-2}
     \end{subfigure}
        \caption{Impact of using splitting rules with different granularity for different demands}
        \label{fig:obs1}
\end{figure}

\begin{figure}[h!]
     \centering
     \begin{subfigure}[b]{0.32\textwidth}
         \centering
         \includegraphics[width=1\textwidth]{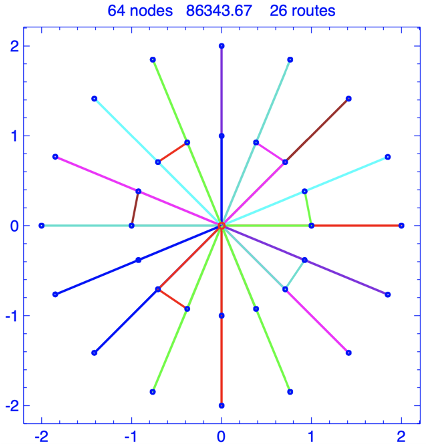}
         \caption{80/40/20/10 rule, \\\textbf{4.98 s}}
         \label{fig:obs2-1}
     \end{subfigure}
     \begin{subfigure}[b]{0.32\textwidth}
         \centering
         \includegraphics[width=\textwidth]{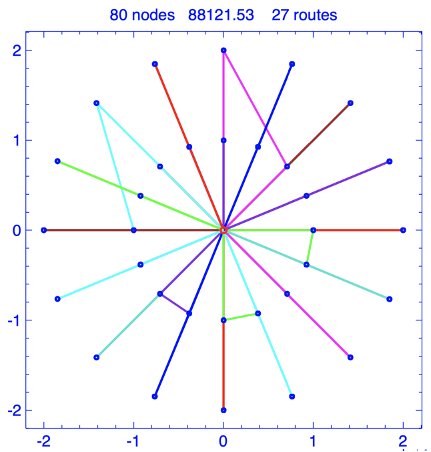}
         \caption{40/20/10 rule,\\ \textbf{5.90 s}.}
         \label{fig:obs2-2}
     \end{subfigure}
     \begin{subfigure}[b]{0.32\textwidth}
         \centering
         \includegraphics[width=\textwidth]{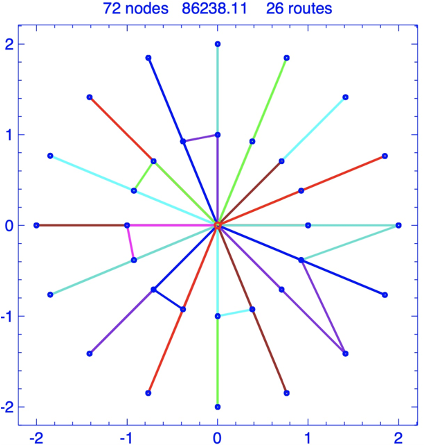}
         \caption{80/40/20/10 and 40/20/10 combined rule, \textbf{5.41 s}}
         \label{fig:obs2-3}
     \end{subfigure}
        \caption{Impact of using splitting rules with different granularity according to the customers' distance to the depot}
        \label{fig:obs2}
\end{figure}

In this section, we present the motivation behind our proposed splitting algorithm through illustrative examples. We study the impact of the splitting rule granularity on the total travel cost and the solution run-time considering customers' demand values and their distances from the depot. Our investigations reveal the benefits of using coarser/finer splitting rules for different customers depending on their demand and location. By a \emph{coarser} splitting rule, we mean one that entails larger splitting portions as opposed to a \emph{finer} splitting rule where the portions are smaller. For instance, for a demand point with $d=200$, the splitting rule $200 = 128+64+8$ is coarser compared to the rule $200= 3\times 64 + 8$.
\\\\\textbf{Impact of coarser rules for customers with higher demands}: 
Using coarser rules for customers with high demands introduces a two-fold benefit: $i$) it reduces the run-time due to a smaller number of resulting customers in the resulting CVRP, $ii$) it often results in a lower total cost of delivery. To capture this effect, we employ an exponential pattern (such as 128/64/32/16/8/4/2) for the \algo{} rule. We illustrate this effect for the $P04\_3070$ instance of the third dataset provided in \cite{data3} as depicted in Fig. \ref{fig:obs1}. In this instance, the vehicle capacity is $200$ and the depot is located at the location $(0,0)$ in Fig. \ref{fig:obs1-1} and Fig. \ref{fig:obs1-2}. Each point in the plane corresponds to a customer. The routes of the vehicles in the obtained solution are denoted by different colors. Figures \ref{fig:obs1-1} and \ref{fig:obs1-2} show the solution of the SDVRP instance considering the $20/10/5/1$ and $128/64/32/16/8/2$ splitting rules, respectively. We observe that the splitting rule of $128/64/32/16/8/4/2$ reduces the solver run-time by an order of $2.5$. The lower run-time is expected since using an exponential pattern in the splitting rule results in splitting the higher demands into a piece with a very high value and most likely fewer pieces in total. This effect can also be observed from the size of the generated CVRP instance. The $20/10/5/1$ rule resulted in a new instance with $855$ customers (nodes), while the $128/64/32/16/8/4/2$ rule generated a network with $572$ customers. Moreover, the coarser splitting rule of $128/64/32/16/8/4/2$ resulted in a cost of $4424.64$ that improves upon the cost of the rule $20/10/5/1$ (proposed by the VRPHAS algorithm) by $1.34\%$. 
\\\\\noindent \textbf{Impact of coarser rules for customers further away from the depot}:
The second motivation behind our proposed \algo{} rule is that using coarser rules for the demands with a larger distance from the depot and splitting closer demands into smaller pieces results in lower solver run-time and possibly a better solution. The intuition behind this observation is that as the distance between a customer and the depot increases, each time visiting that customer would significantly increase the cost. Therefore, it is desirable to satisfy the demand of further customers within the fewest possible number of visits and that is only possible by applying coarser splitting rules to those nodes. 
Moreover, by splitting closer-to-depot customer demands into smaller pieces, we provide the opportunity for the vehicles returning to the depot to utilize their remaining capacity to fulfill smaller demands and, therefore, minimize the fraction of their unused capacity. It is important to note that all of these intuitive statements are based on the fact that after splitting demands and generating a new network with a larger number of customers, we use the VRPH solver and basically solve the generated instance by a number of heuristic and metaheuristic methods. Figures \ref{fig:obs2-1}, \ref{fig:obs2-2}, and \ref{fig:obs2-3} illustrate the above observation for the $SD6$ instance from \cite{golden7}. The best-known solution for this instance is $830.86$. In Fig. \ref{fig:obs2-1} and \ref{fig:obs2-2}, the results of using the $80/40/20/10$ and $40/20/10$ rules are shown. Although these two rules alone provide acceptable results with the costs of $863.43$ and $881.21$, respectively, we observe that by using a combination of the two rules as shown in Fig. \ref{fig:obs2-3}, we can achieve lower cost of $862.38$. In this experiment, the $32$ customers are clustered into two categories of size $16$. For the $16$ customers closer to the depot, we used the splitting rule $40/20/10$, and the coarser splitting rule $80/40/20/10$ is applied to the remaining $16$ customers located farthest from the depot. 

\subsubsection{\algo{} Algorithm}
The above observations highlight the significance of applying an adaptive splitting rule for different customers instead of using a fixed splitting rule for all customers, as in VRPHAS. Motivated by these observations, we incorporate the following factors into \algo{}:

\begin{itemize}
    \item \textbf{Vehicle capacity:} 
    As implied by \cite{base}, it makes sense to look for the splitting rule as a function of the vehicle capacity. This not only presents a structured representation with vehicle capacity as the kernel allowing for better generalization but also leads the policy towards leaving the lowest portion of the vehicles' capacities unused and consequently increases the vehicles' utilization.
    
    \item \textbf{Customer location:} We believe one potential improvement to the VRPHAS splitting rule is to personalize the rule for each customer considering its distance from the depot. We observe that if the customers that are located further from the depot are visited by one (or very few) vehicle(s), the total travel cost is lower. Otherwise, if a long distance needs to be traversed multiple times, the cost will be negatively impacted. This can be avoided by coarsening the splitting rule for further customers. 
    In \algo{}, the impact of customers' locations is addressed by partitioning the customers into a number of clusters based on their distance from the depot. We then use a different splitting rule for each cluster. A coarser splitting rule is used for clusters of customers further from the depot while finer rules are applied to the clusters closer to the depot.
    
    \item \textbf{Customer demand:} Another important factor in the adaptation of the splitting rule is the demand of customers. Depending on how large the demand is, the splitting rule can be applied in full or partially to result in an appropriate number of additional demand points with appropriate demand values. In order to create the exponential pattern in the splitting rule, we propose to decompose the customers' demands based on the different powers of prime numbers. For instance, the demand of a customer with $d = 199$ served by vehicles with capacity $Q= 200$ can be decomposed as $d = 2^7 + 2^6 + 2^2 + 2^1+ 1$ if $p=2$, and $d = 2\times 3^4 + 3^3 + 3^2 +1$ if $p=3$. 
    
\end{itemize}    



\begin{figure}[t]
\begin{center}
\begin{minipage}{1\textwidth}
    \centering
    \includegraphics[width=1\textwidth]{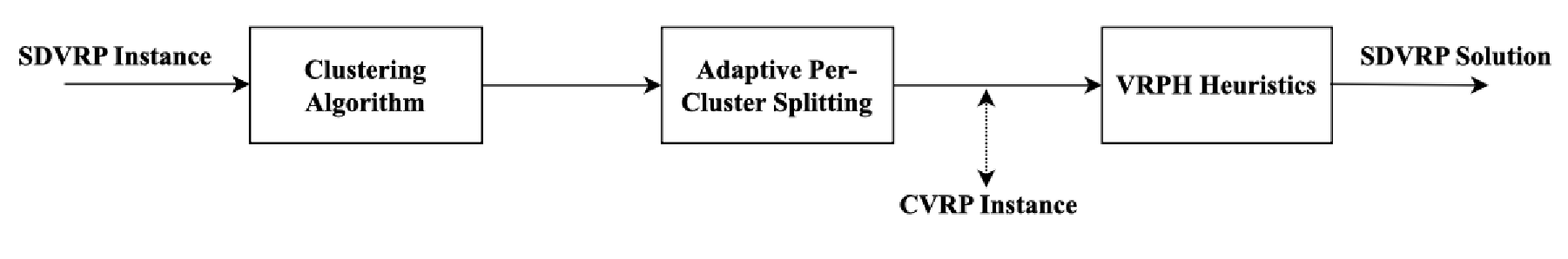}
    \caption{\algo{} framework}
    \label{fig:frame}
\end{minipage}
\end{center}
\end{figure}

Leveraging the above factors, the \algo{} solution framework is shown in Fig. \ref{fig:frame}.
First, PASA employs a clustering algorithm to partition the customers into multiple levels. In our experiments, we used a simple distance-based clustering as follows:
\begin{align}
    label(v) = \ell, \quad \frac{\ell-1}{L}{dist}_{max}  < dist(v, 0) \leq \frac{\ell}{L}{dist}_{max}, \quad \ell\in\{1, \ldots, L\}
	\label{cluslabel}
\end{align}
where $L$ is the number of levels, and ${dist}_{max}$ is the maximum customer-depot distance among all customers, i.e, ${dist}_{max} = max_v\ dist(v, 0)$. As a result of this clustering rule, the customers are separated with $L$ uniformly-spaced rings around the depot. Second, for the customers at each level, the same splitting rule is applied. After all the demands are split a CVRP instance is generated. Finally, the VRPH solver is used to solve the resulting CVRP instance. 

The pseudo-code for \algo{} is given in Algorithm \ref{RSR}. The algorithm starts by taking as input the SDVRP instance as a graph $G = (V, E)$, the vehicle capacity $Q$, the vector of customer-depot distances denoted by $\Vec{dist}$ where $\Vec{dist}(v) = dist(v,0)$, and the vector of customer demands $\Vec{D}$. Next, it computes $d$ the greatest common divisor ($gcd$) of the demands vector and the vehicle capacity, which will be used as a parameter in the adaptive splitting rule. In fact, the vehicles may carry goods only in quantities that are multiples of $d$ units. Next, the demand vector is scaled down by $d$ and averaged over the customers to find $\mu$ that is used in finding $s_{max}$ to determine the largest quantity of goods that is demanded by the customers in the equivalent CVRP model. 

The algorithm takes into account $L$ different levels for the distance-based clustering as provided in \eqref{cluslabel}. Then for each level $i$, the splitting rule $\Vec{s}_i$ is determined using the value $s_{max}$. Next, the clustering of the customers is made according to equation \eqref{cluslabel}. The \emph{split} function substitutes each node (customer) of the graph $G$ with new customers according to the rule $\Vec{s}_i$ to gradually form the graph $G'$. Note that the splitting rule ensures that the demands further away from the depot are split with coarser rules. Once the graph $G'$ is completed, the CVRP solver will solve the generated new problem.
We experimented with different values of the prime number $p$ and observed that $p=2$ has the best performance overall in the current version of \algo{}.

For example, consider the instance $SD6$ from SET-1 as shown in Fig.~\ref{fig:obs2}. We will have $d = gcd(\Vec{D}, Q)$, i.e.,  $d = gcd(60,  90, 10) = 10$. The simple distance-based clustering in equation~\ref{cluslabel} with $L=2$ results in $2$ levels, where $l=1$ covers the customers that are furthest from the depot and  $l=2$ covers the customers that are closest to the depot. We will have $\mu = 7.5$ and $s_{max} = 3$. Therefore, for the first level, we have $\vec{s_1} = \{10, 20, 40, 80\}$ and for $\vec{s_2} = \{10, 20, 40\}$.
\\\\\noindent\textbf{VRPH solver}:
Once the \emph{a priori} splitting rule is applied to an SDVRP problem, any commercial solver can be used to solve the resulting CVRP instance. In this paper, we use the VRPH solver to solve CVRP instances. VRPH is a publicly available solver with a provable record of generating high-quality solutions. It takes as input a CVRP instance written to a file with a format similar to that of the Traveling Salesman Problem Library (TSPLIB) files and prints the solution to an output file. VRPH implements an open-source library of several local search heuristics for generating and improving feasible solutions to the CVRP instances. We refer the reader to \citep{osti_979334} for a detailed description of the heuristics and the modular structure of the VRPH software. In this paper, we use VRPH as a standalone solver with the parameters set at their default values.

\begin{algorithm}[t]
 \caption{\emph{A priori} Adaptive Splitting Algorithm (\textit{\algo{}})}
 \label{RSR}
 \begin{algorithmic}[1]
 \renewcommand{\algorithmicrequire}{\textbf{Input:}}
 \renewcommand{\algorithmicensure}{\textbf{Output:}}
 \REQUIRE $G = (V, E), Q, \Vec{dist}, \Vec{D}, p$
 \ENSURE $G' = (V', E')$
 \\ 
 \STATE Initialize $L, d \xleftarrow{} gcd(Q, \Vec{D})$, $\mu \xleftarrow{} avg(\Vec{D/d}), s_{max} \xleftarrow{}[log_p^\mu]$
 \STATE \textbf{for} $i : 1..L$
 \STATE \quad $ \Vec{s_i} \xleftarrow{} d*(exp_p\{0 ..  s_{max}-i+1\})$
 \STATE \textbf{end for}
 \STATE $\Vec{label} = clustering(V)$ \qquad \qquad \qquad \qquad \qquad \qquad \qquad \qquad \qquad \qquad//{$\Vec{label}(v) \in \{1,...,L\}$}
 \REPEAT
 \STATE $V = V \setminus \{v\}$
 \STATE $i\xleftarrow{} \Vec{label}(v)$ 
 \STATE $V' = V' \cup{}split(v, \Vec{s_i})$
 \UNTIL {$V = \emptyset$}
 \RETURN $G' = (V', E')$
 \end{algorithmic} 
 \end{algorithm}

%% file: evaluation.tex
\section{Performance Evaluation}
\label{sec:evaluation}
In this section, we evaluate the performance of \algo{} through extensive numerical simulations. We benchmark the performance of \algo{} against three baseline strategies, using instances from datasets that are widely used in the literature for the comparison of various vehicle routing solutions. Table \ref{data} presents the summary of the evaluation datasets used throughout this section. Further details about the instances of each data set are provided in Section \ref{evaluation:data}.

\begin{table}[t]
\centering
\caption{Evaluation datasets}
\begin{tabular}{ c|c|c|c|c }
 \hline
 Dataset & Number of instances  &  Number of customers (N) & Vehicle Capacity (Q) & Customers' demands\\
 \hline
 SET-1 & 21 & $[8,288]$ & 100 & $\{60,90\}$\\
 \hline
 SET-2 & 14 & $\{50, 75, 100\}$ & 160 & randomly from $[aQ, bQ]^1$ \\
 \hline
 SET-3 & 42 & $[50,199]$ & $[140,200]$ & randomly from $[aQ, bQ]^1$\\
 \hline
 SET-4& 11  & $[21,100]$ & $[112,8000]$ & No pattern\\
 \hline
\end{tabular}
\\\footnotesize{$^1$ $(a,b) \in \{(0.01, 0.1), (0.1, 0.3), (0.1, 0.5), (0.1, 0.9), (0.3, 0.7), (0.7, 0.9)\}$, corresponding to six different cases}
\label{data}
\end{table}

\subsection{Benchmarking Instances}
\label{evaluation:data}
We consider $4$ benchmarking sets that include $88$ instances overall as presented in Table~\ref{data}. SET-$1$ \citep{golden7} contains $21$ instances, each with customers uniformly distributed on the perimeter of concentric circles centered at the depot. The instances are sorted by the number of customers and the number of concentric circles. 
SET-$2$ \citep{data2} and SET-$3$ \citep{data3} instances use similar customer demand profiles but vary in the number of customers and vehicle capacity. These demands are randomly sampled from the range $[aQ, bQ]$ with $(a, b)$ chosen by six different scenarios as in Table~\ref{data}. SET-$2$ contains $14$ and SET-$3$ contains $42$ instances, $6$ of which are repeated. Therefore, we test only $36$ of them which are not redundant. The coordinates of the customers in SET-$2$ are randomly generated using the coordinates of eil51, eil76, and eil101 from TSPLIB which are also used in SET-$4$ \citep{data4}. SET-$4$ includes $11$ instances where no specific rule is preserved throughout for generating the customer demands.

\subsection{Benchmarking Solutions}
\label{evaluation:benchmark}
We benchmark the performance of the \algo{} algorithm against the following baseline methods:
\begin{itemize}
    \item No-splitting (CVRP): The baseline method that treats an SDVRP instance as a CVRP instance and runs the VRPH \citep{VRPH} solver to generate solutions. 
    \item VRPHAS: The method based on \emph{a priori} splitting that is proposed by \cite{base}.
    \item ILS-MIP: The metaheuristic method proposed by \cite{ILS-MIP}.
\end{itemize}

We note that when computing the optimality gap corresponding to each method for each experimental instance, we take the best solution known in the literature as the reference solution. Further, we note that in ILS-MIP, the travel cost between each pair of customers is assumed to be their rounded Euclidean distance except for SET-1 instances where the exact Euclidean distance is taken as the cost. Hence, we compare the results of ILS-MIP and \algo{} only for SET-1. For the rest of the datasets, we compare the performance of \algo{} against the No-splitting and the VRPHS methods.

\subsection{Metrics and Setup}
We consider the optimality gap and the solver run time as metrics for performance comparison between the above strategies. The optimality gap which indicates the deviation from the best-known solution is defined as:
$$gap = \frac{obj-\text{best-known solution}}{\text{best-known solution}}*100$$
where $obj$ is the objective value of the considered method. We run both VRPHAS and \algo{} algorithms on the same PC with an Intel Xeon processor at $3.2$ GHz and $16$ GB of main memory. It is important to note that the gap results we obtained for the VRPHAS method do not match the values presented in \cite{base} for some instances. However, since we need to compare both the run time and the gap results for the VRPHAS and \algo{} methods on similar hardware, we reflect the results of our run for VRPHAS. Moreover, we report the computation time of ILS-MIP as presented in \cite{ILS-MIP}, including the time values $\text{time}$ and $\text{time}_{\text{best}}$ corresponding to the time required for the proposed iterative algorithm to terminate and the time at which the best solution of ILS-MIP is achieved, respectively. We reflect both these values in Table 2 under the time column for ILS-MIP. A time limit (TL) is considered for the algorithm termination that is assumed to be equal to $1349$s by \cite{ILS-MIP}. It is important to note that the ILS experiments are conducted on a PC with an Intel Core i7-8700 3.2 GHz processor and 32 GB of memory which is more powerful than our PC. Therefore, we can expect that ILS-MIP performs even worse in terms of computation time compared to \algo{} if both were run on the same system. 

\subsection{Numerical Results}
Table \ref{tab:set-1} presents the performance of \algo{} against VRPHAS, ILS-MIP, and the no-splitting strategies for the instances of SET-1. First, we observe that the no-splitting method results in very low-quality solutions as the resulting optimality gap obtained by this method is very high compared to the VRPHAS, \algo{}, and ILS-MIP heuristics. Moreover, we observe that while the ILS-MIP strategy outperforms the \algo{} and VRPHAS solution approaches in terms of optimality gap for almost all instances of SET-1, it requires very large computation times. The average run-time for ILS-MIP is $992.38s$ as opposed to the average run-time of $86.14s$ and $30.56$s corresponding to the VRPHAS and \algo{} solution approaches, respectively. This huge time complexity hinders the practical application of the ILS-MIP approach in real-world problems, especially as the problem size increases. Figures \ref{fig:set1-2} and \ref{fig:set1-3} illustrate the per-instance optimality gap and run time of our proposed splitting rule against the fixed splitting rule introduced in VRPHAS. Our proposed method \algo{} outperforms the VRPHAS in terms of time complexity for all instances, and it also improves upon the average optimality gap of VRPHAS.

Table \ref{tab:set-2} shows the results of VRPHAS, \algo{}, and the no-splitting methodologies for the instances of SET-2. According to Table \ref{tab:set-2}, the no-splitting approach results in a higher gap compared to both the VRPHAS and \algo{} rules. Furthermore, our proposed \algo{} strategy outperforms the VRPHAS method in terms of the average optimality gap, as well as run-time. Moreover, the per-instance results depicted in Fig. \ref{fig:set2-2} and Fig. \ref{fig:set2-3} indicate that the \algo{} strategy improves upon the optimality gap and solver run-time of the VRPHAS method for almost all instances of SET-2. Similarly, the comparison of results for SET-3 and SET-4 provided in Table \ref{tab:set-3} and Table \ref{tab:set-4} show that our proposed splitting algorithm again improves both average optimality gap and run-time of the VRPHAS splitting mechanism and, therefore, it can effectively generate higher-quality solutions, faster than VRPHAS. The improvement in the run-time is mainly owing to the rule granularity consideration introduced in \algo{} as discussed in Section \ref{sec:solution}, which results in smaller instances to be solved by the CVRP solver compared to the VRPHAS method. 
\input{resultsSet1}
\input{resultsSet2-4Short}

%% file: resultsSet1.tex
\begin{table}[]
    \centering
    \label{table:SET-1}
    \caption{Performance of VRPHAS, \algo{} and ILS-MIP on SET-1}
\begin{tabular}{cc|cc|ccc|cc|c}
 \hline
 \multicolumn{1}{c}{Instance} & \multicolumn{1}{c}{Best-known} &  \multicolumn{2}{c}{VRPHAS} &  \multicolumn{3}{c}{\algo{}}  & \multicolumn{2}{c}{{ILS-MIP}}  &\multicolumn{1}{c}{No Splitting}\\
  & Solution & gap ($\%$) & time(s) & cost & gap ($\%$) & time (s) &gap  & $\text{time}$ - $ \text{time}_{\text{best}}$ (s) & gap ($\%$) \\
 \hline
  SD1 & 228.28  & 0.00 & 0.51 & 228.28  & 0.00  & 1.60  &\red{0.00} & \red{25.10} - \blue{0.00} & 5.13 \\
  
  SD2 & 708.28 & 0.00 & 1.23 & 708.28 & 0.00 & 2.85  &\red{0.00} & \red{47.51} - \blue{0.00} & 12.94 \\
  
  SD3 & 430.40   & 0.04 & 1.25 & 430.58  & 0.04  & 2.57 &\red{0.05}& \red{37.68} - \blue{0.00} & 11.51\\
  
  SD4 & 630.62  & 0.07 &  2.35 & 631.04  & 0.07  & 3.57 &\red{0.07} & \red{75.45} - \blue{0.00} & 14.17\\
  
  SD5 & 1389.94  & 1.22 & 3.91 & 1390.57  & 0.05  & 5.66 &\red{0.05}& \red{216.69} - \blue{0.80} & 15.10\\
  
  SD6 & 830.86  & 0.05 & 3.85 & 831.24 & 0.05  & 5.09 &\red{0.04}& \red{202.60} - \blue{92.42} & 15.47\\
  
  SD7 & 3640.00  & 0.55 & 5.47 & 3640.00 & 0.00  & 7.79  &\red{0.00}& \red{TL} - \blue{0.00} & 20.87\\
  
  SD8 & 5068.28  & 0.00 & 7.64 & 5068.28 & 0.00 & 10.43  &\red{0.00}& \red{TL} - \blue{0.00} & 23.11\\
  
  SD9 & 2042.88  & 0.72 & 7.50 & 2051.06 & 0.40  & 8.97  &\red{0.07} & \red{TL} - \blue{39187} & 17.47 \\
  
  SD10 & 2683.73 & 0.95 & 13.73 & 2704.88 & 0.79  & 13.68   &\red{0.38} & \red{TL} - \blue{960.42} & 19.23\\
  
  SD11 & 13280.00  & 0.15 &20.59 & 13480.00 & 1.50  & 18.43   &\red{0.00} & \red{TL} - \blue{0.00} & 26.50  \\
  
  SD12 & 7213.62  &  0.73 & 20.89 & 7238.88 & 0.34  & 20.99   &\red{0.09} & \red{TL} - \blue{1223.31} & 21.99\\
  
  SD13 & 10105.86  & 0.05 & 25.90 & 10110.58 & 0.05  & 22.05   &\red{0.05} & \red{TL} - \blue{1091.84} & 23.49\\
  
  SD14 & 10717.53 & 0.70 & 42.57 & 10804.84 & 0.81  & 32.03   &\red{0.22} & \red{TL} - \blue{967.30} & 23.16\\
  
  SD15 & 15094.48  & 0.93 & 59.43 & 15232.60 & 0.91  & 47.32  &\red{0.37} & \red{TL} - \blue{318.44} & 24.01\\
  
  SD16 & 3379.33  & 5.50 & 54.88 & 3508.16 & 3.81  & 30.16   &\red{0.11} & \red{TL} - \blue{0.01} & 27.83\\
  
  SD17 & 26493.56  & 0.25 & 66.98 & 26962.64 & 1.56  & 50.14   &\red{ 0.1} & \red{TL} - \blue{1328.05} & 26.82\\
  
  SD18 & 14202.53  &0.69 &72.04 & 14278.70 & 0.54 & 54.75   &\red{0.46}& \red{TL} - \blue{1347.05} & 23.92\\
  
  SD19 & 19995.69  & 1.00 & 101.05 & 20197.04 & 1.00  & 70.14   &\red{0.64} & \red{TL} - \blue{1328.17} & 24.82\\
  
  SD20 & 39635.51  &0.17 & 162.90 & 40236.66 & 1.51 & 104.51   &\red{0.25} & \red{TL} - \blue{1328.44} & 27.15\\
  
  SD21 & 11271.06  & 3.41 &240.36 & 11576.96 & 2.71  & 129.31  &\red{0.2} & \red{TL} - \blue{0.07} & 27.76 \\
  \hline
  Average & & 0.82 & 86.14 & & 0.76 & 30.56  & \red{0.15} &\red{992.38} - \blue{494.17} & 20.60\\
 \hline
 \label{tab:set-1}
\end{tabular}
\end{table}

\begin{figure}[htb]
\begin{center}
\begin{minipage}[h]{0.48\textwidth}
  \includegraphics[width=1\textwidth]{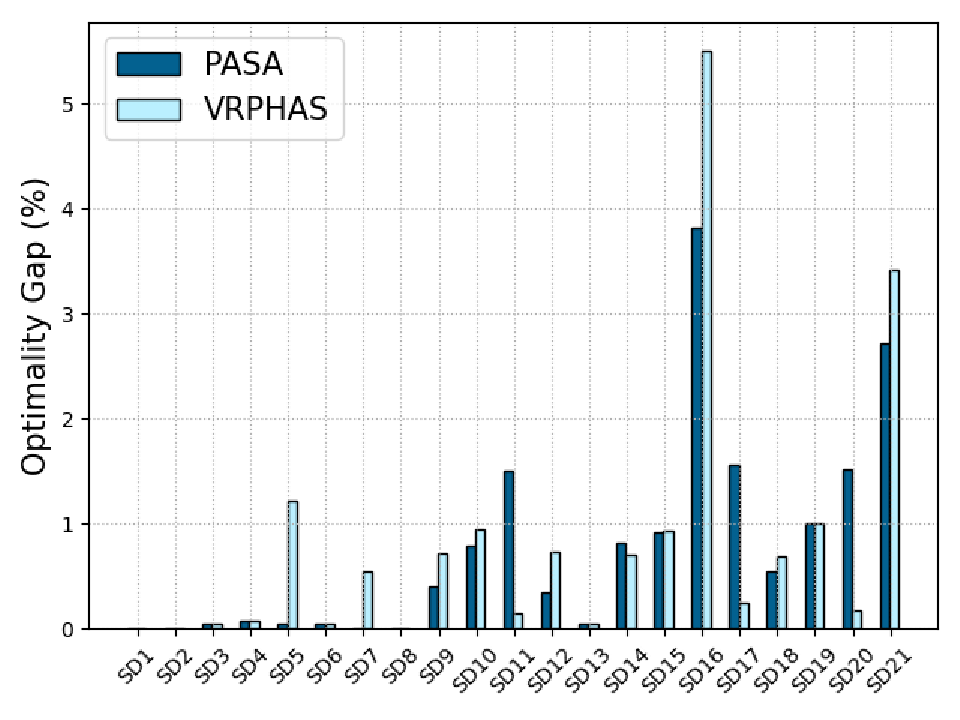}
  \caption{Per-Instance Gap of SET-1}
    \label{fig:set1-2}
\end{minipage}
\begin{minipage}[h]{0.48\textwidth}
  \includegraphics[width=1\textwidth]{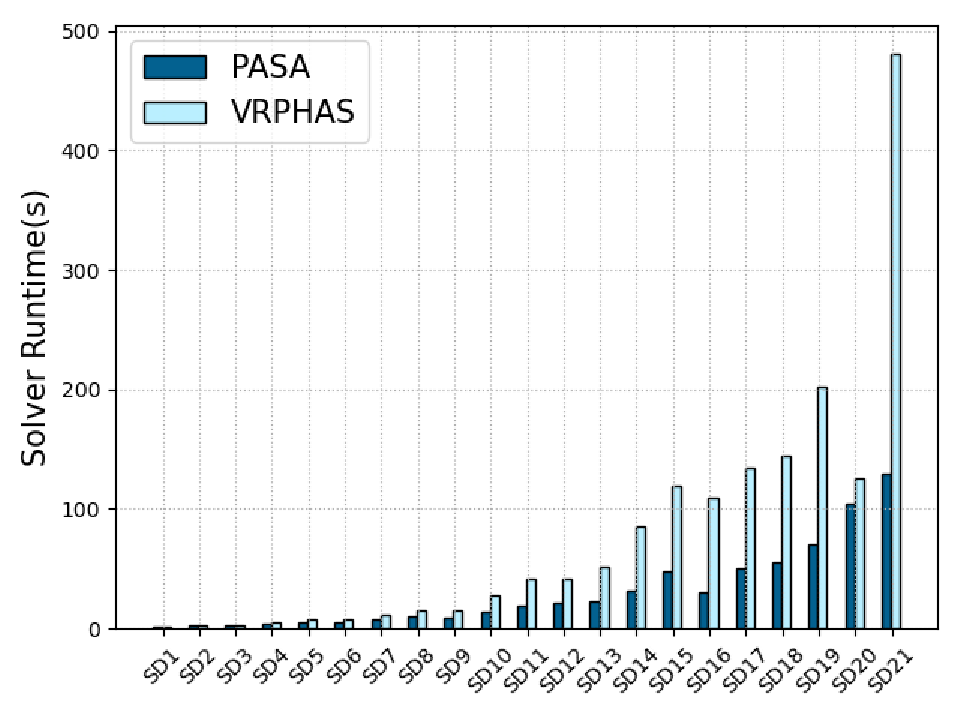}
  \caption{Per-Instance Solver Run-time of SET-1}
    \label{fig:set1-3}
\end{minipage}
\end{center}
\end{figure}
\newpage

%% file: resultsSet2-4Short.tex
\begin{table}[t]
\caption{Performance of VRPHAS, and \algo{} on SET-2}
\centering
\begin{tabular}{cc|cc|ccc|c}
 \hline
 \multicolumn{1}{c}{Instance} & \multicolumn{1}{c}{Best-known} &  \multicolumn{2}{c}{VRPHAS} &  \multicolumn{3}{c}{\algo{}} &  \multicolumn{1}{c}{No Splitting}  \\
  & Solution & gap ($\%$) & time (s) & cost & gap ($\%$) & time (s)  & gap ($\%$) \\
 \hline
S51D1 & 459.50  & 0.00 & 10.98 & 459.50 & 0.00 & 7.92 & 0.00 \\ 

S51D2 &  709.29 & 0.77 & 15.04 & 713.80  & 0.63  & 11.22 & 1.71 \\ 

S51D3 & 948.06 & 1.88  & 20.28  & 959.92  & 1.25 & 13.04 &  3.19 \\ 

S51D4 & 1562.01 & 1.94 & 28.24 & 1583.88 & 1.40 & 22.94 & 7.32 \\

S51D5 & 1333.67 & 1.03  & 30.68 & 1343.109  & 0.71  & 20.81 & 8.89\\ 

S51D6 & 2169.10 & 1.86 & 43.81 &  2211.90  & 1.97 & 23.37 & 10.75\\ 

S76D1 & 598.94 &  0.00 & 23.18 & 598.94 & 0.00 & 14.55 & 0.78 \\ 

S76D2 & 1087.40 & 1.82  & 32.48  & 1104.88 & 1.61 & 22.43 & 2.79 \\

S76D3 & 1427.86 & 1.93 & 46.99 & 1436.99 & 0.64 & 28.05  & 2.52 \\

S76D4 &  2079.76  & 2.51 & 55.89  & 2120.16 & 1.94 & 40.09  & 5.41\\

S101D1 & 726.59 & 1.23 & 34.03 & 729.72 & 0.43 & 42.31 & 1.14\\ 

S101D1 & 726.59 & 1.23 & 34.03 & 729.72 & 0.43 & 42.31 & 1.14\\ 

S101D3 &  1874.81 & 1.85 & 70.64 &  1905.43 & 1.63 & 41.05 & 4.13\\

S101D5 & 2791.22 & 1.61 & 95.71 & 2844.51 & 1.91 & 62.54 & 11.59
\\\hline
Average &   & 1.45 & 40.02 &  & 1.13 & 28.18 & 4.42\\
\hline
\label{tab:set-2}
\end{tabular}
\end{table}

\begin{figure}[htb]
\begin{center}
\begin{minipage}[h]{0.48\textwidth}
  \includegraphics[width=1\textwidth]{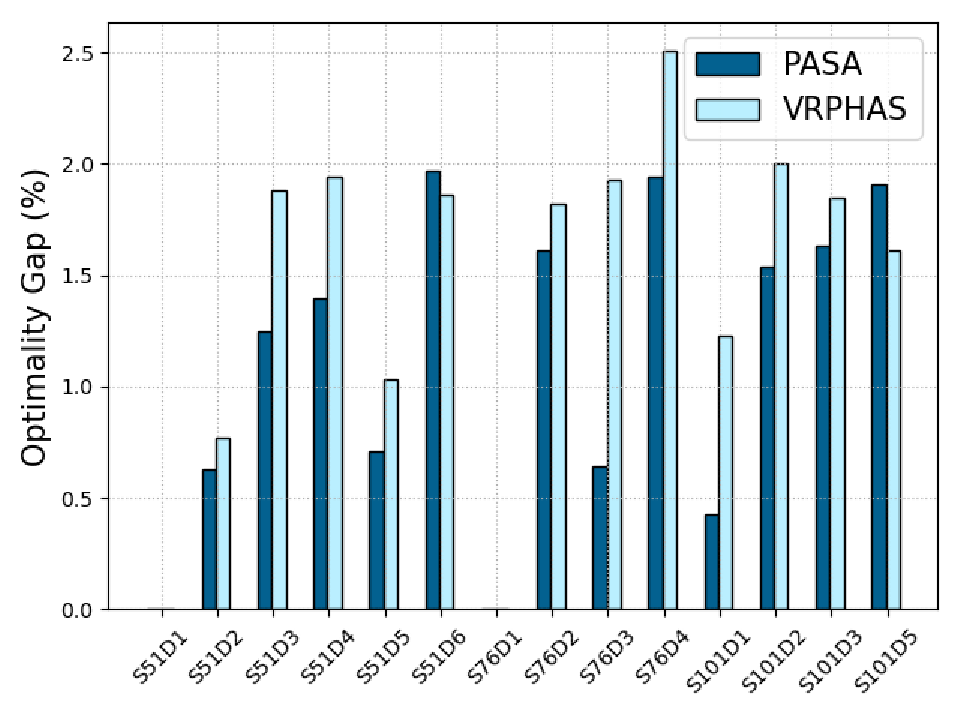}
  \caption{Per-Instance Gap of SET-2}
    \label{fig:set2-2}
\end{minipage}
\begin{minipage}[h]{0.48\textwidth}
  \includegraphics[width=1\textwidth]{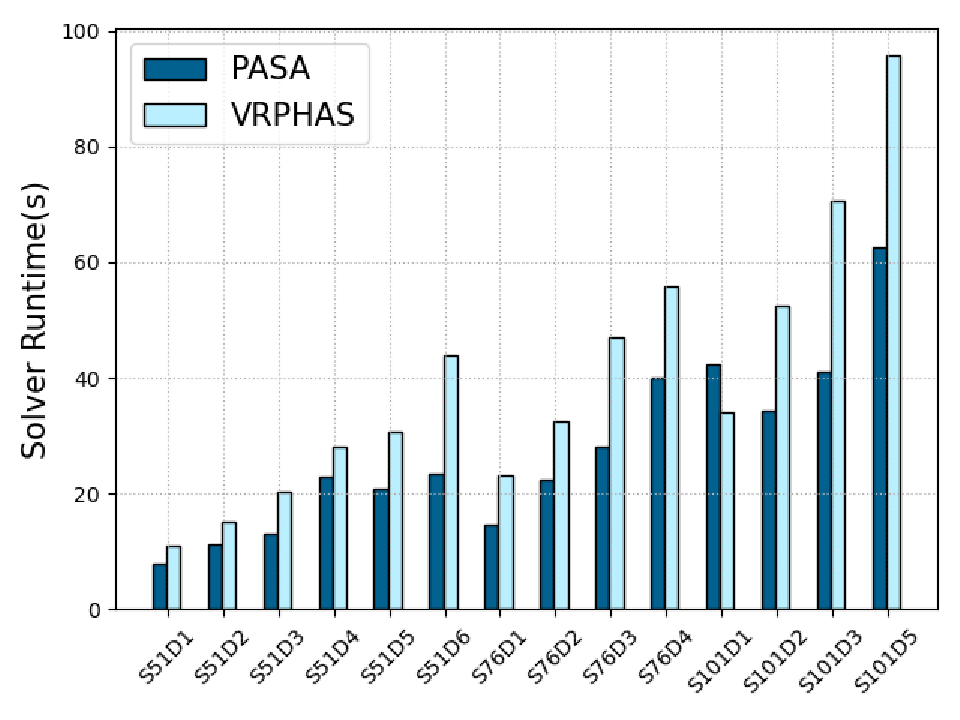}
  \caption{Per-Instance Solver Run-time of SET-2}
    \label{fig:set2-3}
\end{minipage}
\end{center}
\end{figure}

\begin{table}[t]
 \caption{Performance of VRPHAS, and \algo{} on SET-3}
    \centering
\begin{tabular}{cc|cc|ccc|c}
 \hline
 \multicolumn{1}{c}{Instance} & \multicolumn{1}{c}{Best-known} &  \multicolumn{2}{c}{VRPHAS} &  \multicolumn{3}{c}{\algo{}} &  \multicolumn{1}{c}{No Splitting} \\
  &Solution& gap ($\%$) & time (s) & cost & gap ($\%$) & time (s)  & gap ($\%$) \\
 \hline
p01\_110  & 459.50  & 0.00 & 9.84 & 459.50  & 0.00  & 6.63 & 0.00\\

p01\_1030 & 757.17 & 1.74 & 17.2 & 770.93 & 1.81 & 11.08 & 3.29\\

p01\_1050 & 1005.75 & 2.56 & 22.52 &  1021.18 & 1.53 &  15.76 & 4.08\\

p01\_1090 & 1488.58 & 1.48 & 32.70  & 1507.14  & 1.24  & 19.61 & 12.75\\ 

p01\_3070 & 1481.71 & 2.23 & 32.84 & 1503.66 & 1.48 & 17.02 & 13.81\\

p01\_7090 & 2156.14 & 2.24 & 49.82 & 2203.93 &  2.22 & 21.41 & 11.41\\ 

p02\_110 &  617.85  & 0.66 & 21.82 & 618.13 & 0.04 &12.08 & 0.65\\

p02\_1030 & 1109.62 & 1.65 & 28.86 & 1125.65 & 1.44 & 21.83 & 2.50\\ 

p02\_1050 & 1502.05   & 1.06 & 22.52 & 1508.98  & 0.46 & 26.21 & 2.72\\

p02\_1090 & 2298.58   & 2.00 & 61.01 & 2334.33 & 1.55 & 41.19 & 9.21\\

p02\_3070 & 2219.97  & 2.23 & 58.80 &  2269.50 & 2.23 & 28.02 & 12.87\\ 

p02\_7090 & 3223.4   & 1.57 & 89.76 & 3305.53 & 2.54 & 40.52 & 12.64\\

p03\_110 &  752.62  & 1.78 & 55.12 & 762.54  & 1.31 & 22.08 & 1.51\\

p03\_1030 & 1458.46   & 1.87 & 37.26 &  1476.86 & 1.26 & 35.67  &2.09\\

p03\_1050 & 1996.76 & 2.54 & 58.40 & 2032.25 & 1.77 & 41.76 & 3.93\\

p03\_1090 & 3085.69   & 2.29 & 143.32 &  3143.69 & 1.88 & 56.56 & 12.17\\ 

p03\_3070 & 2989.3   & 2.19 & 136.10 & 3034.28 & 1.50 & 40.02 & 16.50\\ 

p03\_7090 & 4387.32  & 1.66 & 186.14 & 4464.65 & 1.76 & 84.48 & 13.72\\ 

p04\_110 &  919.17  & 1.35 & 116.08 & 925.19  & 0.65 & 33.83 & 0.84 \\ 

p04\_1030 & 2016.97  &  2.71 & 152.44 &  2056.71 & 1.97 & 55.37 & 2.97\\
 
p04\_1050 & 2849.66  & 2.40 & 196.82 & 2889.62 &  1.40 & 72.14   &  5.09\\ 

p04\_1090 & 4545.46 & 2.44 & 310.84 & 4624.18 & 1.73 & 112.41 & 14.72\\

p04\_3070 & 4334.71   & 2.87 & 272.44  & 4407.02 & 1.67 & 99.65 & 19.64\\ 

p04\_7090 & 6395.41   & 1.94 &  469.62 & 6555.47  & 2.5 & 82.07 & 15.09\\ 

p05\_110 &  1074.18  & 2.74 & 186.66 & 1086.27 & 1.12 & 66.5 & 1.42 \\

p05\_1030 & 2478.4   & 1.83 &  241.44 & 2515.45 & 1.49 & 82.72  & 2.46\\ 

p05\_1050 & 3471.41 & 1.54 & 311.94 & 3510.99 & 1.14 & 87.56  & 4.01\\ 

p05\_1090 & 5521.57   & 2.02 &  561.54 & 5619.52 & 1.77 & 145.02 & 12.67\\ 

p05\_3070 & 5409.76   &  2.41 & 383.28 & 5412.85 & 0.06 & 142.81 & 15.57\\ 

p05\_7090 & 8192.03  & 1.96 & 884.68 & 8332.36 & 1.71 & 255.05 & 17.28\\

p11\_110 & 1031.11 & 2.11 & 74.12 &1044.44 & 1.29  & 20.20 & 1.55\\

p11\_1030 & 2881.8   & 2.89 & 94.06 &  2923.86 & 1.46  & 40.64 & 1.95\\ 

p11\_1050 & 4219.01   &  2.29 & 108.54  & 4268.79  & 1.18 & 55.27 & 2.71\\

p11\_1090 & 6854.09   & 3.27 & 188.30  &  6994.56 & 2.05 & 64.06 & 13.22\\ 

p11\_3070 & 6671.04   &  1.44 &  162.36 & 6759.99 & 1.33 & 59.01 & 15.79\\ 

p11\_7090 & 10204.81  &  1.99 & 231.60 & 10406.25 &  1.97 & 101.02 & 19.93\\ 
\hline
Average & & 1.96 & 166.94 & & 1.45 & 33.38 & 8.08 \\\hline
\label{tab:set-3}
\end{tabular}
\end{table}

\begin{figure}[htb]
\begin{center}
\begin{minipage}[h]{0.99\textwidth}
  \includegraphics[width=1\textwidth]{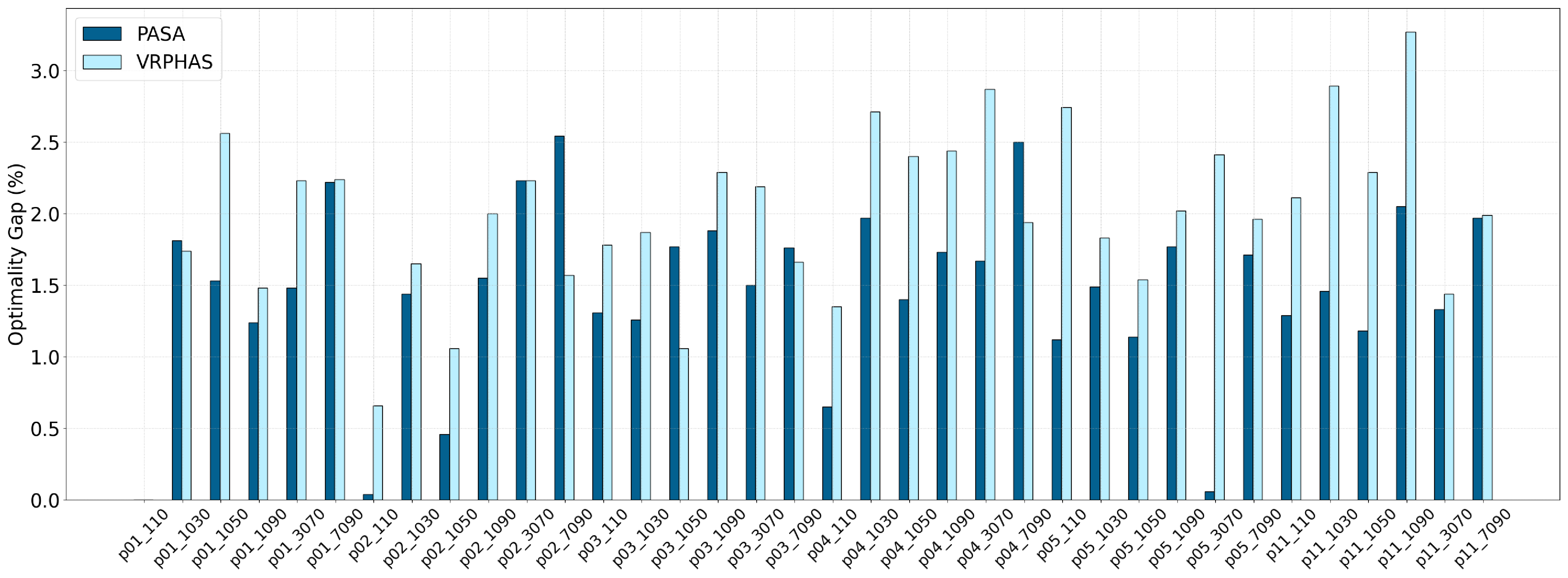}
  \caption{Per-Instance Gap of SET-3}
    \label{fig:set3-2}
\end{minipage}
\begin{minipage}[h]{0.99\textwidth}
  \includegraphics[width=1\textwidth]{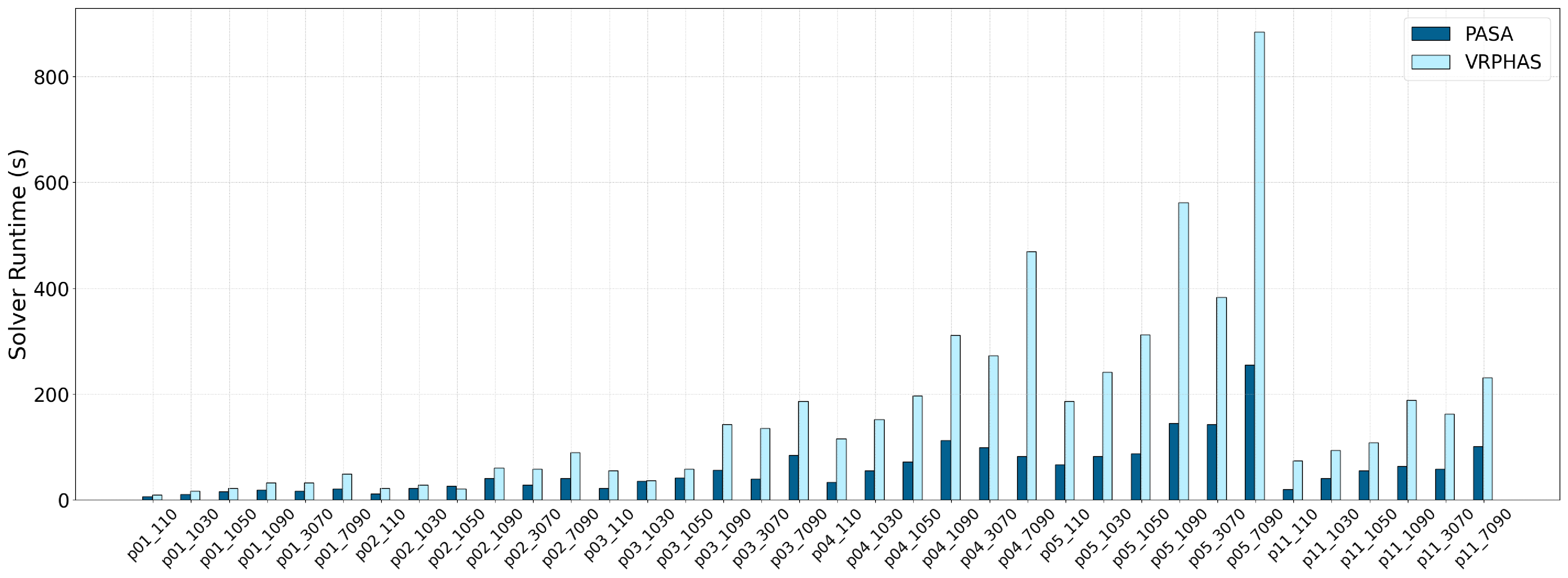}
  \caption{Per-Instance Solver Run-time of SET-3}
    \label{fig:set3-3}
\end{minipage}
\end{center}
\end{figure}

\begin{table}[t]
\caption{Performance of VRPHAS, and \algo{} on SET-4}
    \centering
\begin{tabular}{ cc|cc|ccc|c }
 \hline
 \multicolumn{1}{c}{Instance} & \multicolumn{1}{c}{Best-known} &  \multicolumn{2}{c}{VRPHAS} &  \multicolumn{3}{c}{\algo{}} & \multicolumn{1}{c}{No splitting}  \\
  & Solution & gap ($\%$) & time (s) & cost & gap ($\%$) & time (s) & gap ($\%$) \\
 \hline
eil22 & 375.28 & 0.00 & 3.28 & 375.28 & 0.00 & 3.11 & 0.00\\

eil23 & 568.56 & 0.00 & 5.92 & 568.56 & 0.00 & 2.92 & 0.00 \\

eil30 & 497.53 & 1.50 & 6.89 & 505.01 & 1.50 & 3.64 & 1.48 \\
 
eil33 & 826.41 & 1.36 & 6.89 & 837.67 & 1.36 & 5.08 & 1.36\\

eil51 & 524.61 & 0.00 & 12.58 & 524.61 & 0.00 & 10.95 & 0.00\\

eilA76 &  823.89 & 0.43  & 24.60 & 830.97 & 0.86 & 19.77 & 1.61\\

eilB76 & 1009.04  & 1.96 & 31.86 & 1025.24 & 1.60 & 18.65 & 1.86\\

eilC76 &  738.67 & 0.52 & 26.78 & 742.49 & 0.55 & 17.48 & 0.64\\

eilD76 & 684.53 & 0.81 & 23.82 & 689.47 & 0.72 & 19.41 & 0.82\\

eilA101 & 812.51 & 2.35 & 55.39 & 828.98 & 2.02 & 23.37 & 1.92\\

eilB101 &  1076.26  &  1.57 & 30.90 & 1087.70 & 1.06 & 23.76 & 1.53\\
 \hline
 Average & & 0.96 & 20.80 & & 0.88 & 13.63 & 1.02\\
 \hline
 \label{tab:set-4}
\end{tabular}
\end{table}

\begin{figure}[htb]
\begin{center}
\begin{minipage}[h]{0.48\textwidth}
  \includegraphics[width=1\textwidth]{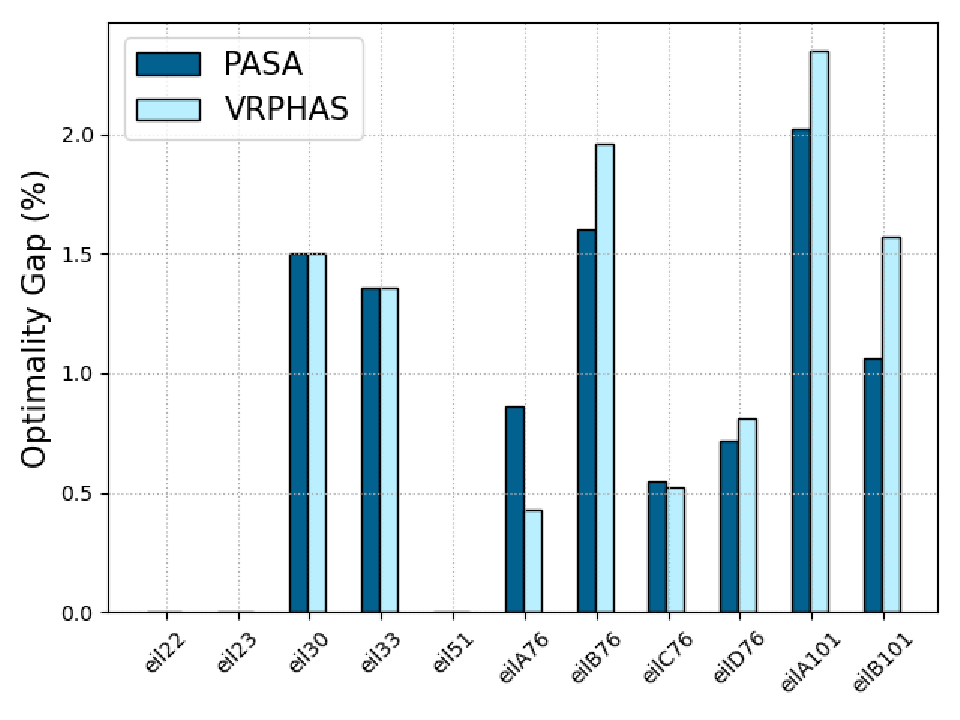}
  \caption{Per-Instance \algo{} Gap of SET-4}
    \label{fig:set4-2}
\end{minipage}
\begin{minipage}[h]{0.48\textwidth}
  \includegraphics[width=1\textwidth]{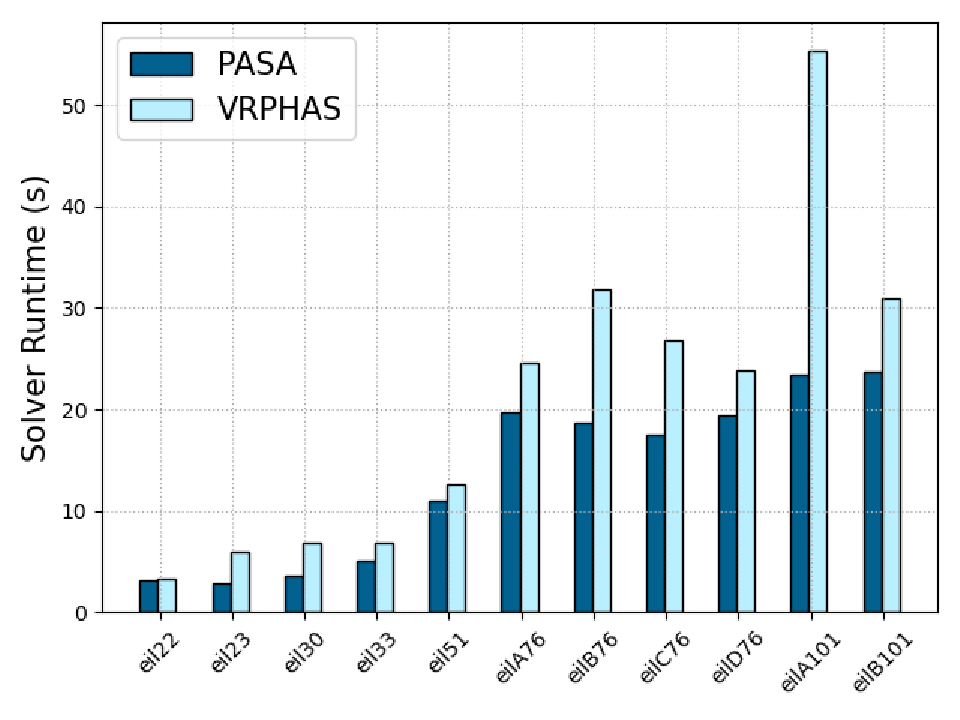}
  \caption{Per-Instance Solver Run-time of SET-4}
    \label{fig:set4-3}
\end{minipage}
\end{center}
\end{figure}